\documentclass{easychair}
\usepackage{makeidx}

\newcommand{\xf}[1]{Figure~\ref{#1}}

\newcommand{\gipc}{{GIPC\index{GIPC}\index{Frameworks!GIPC}}}

\newcommand{\gee}{{GEE\index{GEE}\index{Frameworks!GEE}}}

\newcommand{\gipsy}{{GIPSY\index{GIPSY}}}

\newcommand{\ripe}{{RIPE\index{RIPE}\index{Frameworks!RIPE}}}

\newcommand{\dfg}{{DFG\index{DFG}}}

\newcommand{\lucid}{{Lucid\index{Lucid}}}
\newcommand{\ilucid}{{Indexical Lucid\index{Indexical Lucid}}}

\newcommand{\flucid}{{Forensic Lucid\index{Forensic Lucid}}}

\newcommand{\jooip}{{JOOIP\index{JOOIP}}}

\newcommand{\marfl}{{MARFL\index{MARFL}}}

\newcommand{\java}{{Java\index{Java}}}

\newcommand{\aspectj}{{AspectJ\index{AspectJ}}}

\newcommand{\tool}[1]{\texttt{#1}\index{Tools!#1}}

\newcommand{\api}[1]{\texttt{#1}\index{API!#1}}

\newcommand{\lucidL}[1]{{$\mathit{Lucid}$}($L$) }

		{}

\def\myvert{\raise 2.27pt \hbox{\vrule depth 0pt height 8pt width 0.2mm}}
\def\myarrow{\hspace*{0.43mm}%
             \raise 2.29pt\hbox{\vrule depth 0pt height 8pt width 0.16mm}%
             \hspace*{-0.32mm}%
             $\longrightarrow$
             \ %
             }

\newcommand
	{\graphviz}
	{Graphviz\index{Graphviz}\index{Tools!Graphviz}}

\newcommand
	{\puredata}
	{PureData\index{PureData}\index{Tools!PureData}}

\makeindex

\begin{document}

\title{The Need to Support of Data Flow Graph Visualization of Forensic Lucid Programs, Forensic
Evidence, and their Evaluation by {\gipsy}}
\titlerunning{The Need of DFGs for {\flucid} Programs in {\gipsy}}

\author{
Serguei A. Mokhov\\
       \affiliation{Concordia University}\\
       \affiliation{Montreal, QC, Canada}\\
       \texttt{mokhov@cse.concordia.ca}
\and
Joey Paquet\\
       \affiliation{Concordia University}\\
       \affiliation{Montreal, QC, Canada}\\
       \texttt{paquet@cse.concordia.ca}
\and
Mourad Debbabi\\
       \affiliation{Concordia University}\\
       \affiliation{Montreal, QC, Canada}\\
       \texttt{debbabi@ciise.concordia.ca}
}

\authorrunning{Mokhov, Paquet, Debbabi}

\maketitle

\begin{abstract}
Lucid programs are data-flow programs and can be visually
represented as data flow graphs (DFGs) and composed visually.
{\flucid}, a {\lucid} dialect, is a language to specify and
reason about cyberforensic cases. It includes the encoding of the
evidence (representing the context of evaluation) and the
crime scene modeling in order to validate claims against the model
and perform event reconstruction, potentially within large swaths
of digital evidence. To aid investigators to model the scene and
evaluate it, instead of typing a {\flucid} program, we propose
to expand the design and implementation of the Lucid DFG programming
onto {\flucid} case modeling and specification to enhance the
usability of the language and the system and its behavior. 
We briefly discuss the related work on visual programming an DFG modeling
in an attempt to define and select one approach or a composition of approaches
for {\flucid} based on various criteria such as previous implementation,
wide use, formal backing in terms of semantics and translation.
In the end, we solicit the readers' constructive, opinions, feedback, comments,
and recommendations within the context of this short discussion.%
\end{abstract}

\section{Overview}

Cyberforensic analysis has to do with automated or semi-automated
processing of, and reasoning about, digital evidence, witness accounts, and other details
from cybercrime incidents (involving computers, but not limited
to them). Analysis is one of the phases in cybercrime investigation
(while the other phases focus on evidence collection, preservation,
chain of custody, information extraction that precede the analysis).
The phases that follow the analysis are formulation of a report
and potential prosecution, typically involving expert witnesses.
There are quite a few techniques, tools (hardware and software),
and methodologies have been developed for the mentioned
phases of the process. A lot of attention has
been paid to the tool development for evidence collection
and preservation; a few tools have been developed to aid
data ``browsing'' on the confiscated storage media, log files, memory, and so
on. A lot less number of tools have been developed for case analysis
of the data (e.g. Sleuthkit), and the existing commercial packages (e.g. Encase or
FTK) are very expensive. Even less so there are case management,
event modeling, and event reconstruction, especially with a solid
formal theoretical base. The first formal approach to the cybercrime
investigation was the finite-state automata (FSA) approach by
Gladyshev et. al~\cite{printer-case,blackmail-case}.
Their approach, however,
is unduly complex to use and to understand for non-theoretical-computer science
or equivalent minded investigators.

The aim of {\flucid} is to alleviate those difficulties,
be sound and complete, expressive and usable, and provide even
further usability improvements with the GUI to do
data-flow graph-based (DFG) programming that allows translation between
DFGs and the {\flucid} code for compilation and evaluation.
In a previous related work a similar solution for {\ilucid} was implemented 
in the General Intensional Programming System ({\gipsy})
already~\cite{yimin04}, but requires additional forensic and imperative
extensions.

The goal of {\flucid}
in the cyberforensic analysis is to be able to express in a program
form the encoding of the evidence, witness stories, and evidential
statements, that can be tested against claims to see if there is
a possible sequence or multiple sequences of events that explain
a given {\em story}. As with the Gladyshev's FSA, it is designed to aid investigators to avoid ad-hoc
conclusions and have them look at the possible explanations the {\flucid}
program ``execution'' would yield and refine the investigation, as
was shown in the works by Gladyshev {\em et al.}~\cite{printer-case,blackmail-case}
where hypothetical investigators failed to analyze all
the ``stories'' and their plausibility before
drawing conclusions. 

In \xf{fig:flucid-cred-data-flow} \cite{self-forensics-jooip-flucid-assl}
is a general design overview of the {\flucid} compilation and evaluation
process involving various components and systems. Of main interest to
this work are the inputs to the compiler -- the {\flucid} fragments
(hierarchical context representing the evidence and witness accounts)
and programs (descriptions of the crime scenes as transition functions)
can come from different sources, including the visual interactive {\dfg}
editor that would be used by the investigators at the top-right corner
of the image. Once the complete evidential knowledge of the case and
the crime scene model are composed, the whole specification is compiled
by the compiler depicted as {\gipc} on the image (General Intensional
Program Compiler). The compiler produces an intermediate version of
the compiled program as an AST and a contextual dictionary of all
identifiers among other things, that evaluation engines (under the
{\gee} component) understand. The proposed {\flucid} engines are designed to use
the traditional eduction, {\aspectj}-based tracing, and probabilistic
model checking with PRISM.

\begin{figure}[htpb]
	\includegraphics[width=\columnwidth]{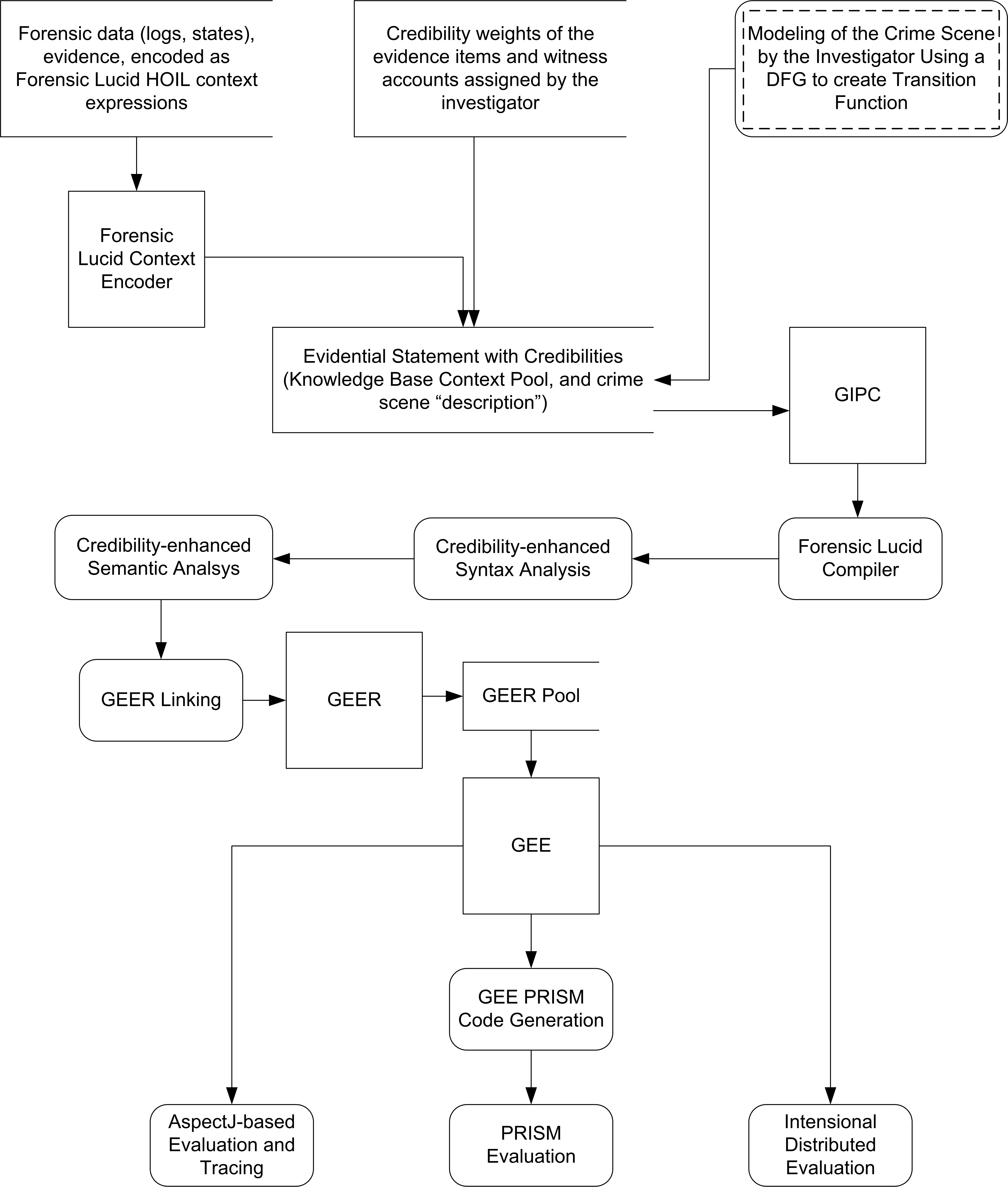}
	\caption{{\flucid} Compilation and Evaluation Flow in {{\gipsy}}}
	\label{fig:flucid-cred-data-flow}
\end{figure}

\section{Related Work}
\label{sect:related-work}

There are a number of items and proposals in graph-based visualization
and the corresponding languages.

In {\gipsy},
our own work in the area includes the theoretical foundation and
initial practical implementation of the DFGs \cite{paquetThesis,yimin04}.
First, Faustini proved that any {\ilucid} program can be represented
as a {\dfg} \cite{denotational-operational-semantics-dataflow}; Paquet
subsequently expanded on this for multidimensional intensional
programs as e.g. shown in \xf{fig:nat42dfg} \cite{paquetThesis}.
Ding further materialized to a good extent Paquet's notion
within the {\gipsy} projects in \cite{yimin04} in 2004 using
\cite{graphviz}'s \tool{lefty}'s GUI and \tool{dot}
languages \cite{dot-language} along with bi-directional translation
between GIPL's or {\ilucid}'s abstract syntax trees (ASTs) to \tool{dot}'s
and back.

Additionally, a part of the proposed related work on
visualization and control of communication patterns and load balancing
idea was to have a ``3D editor'' within {\ripe}'s \api{DemandMonitor}
that will render in 3D space the current communication patterns of a GIPSY
program in execution or replay it back and allow the user visually to
redistribute demands if they go off balance between workers. A kind of
virtual 3D remote control with a mini expert system, an input from which can be used to teach
the planning, caching, and load-balancing algorithms to perform efficiently
next time a similar GIPSY application is run as was proposed in \cite{mokhovmcthesis05}.
Related work by several researchers on visualization of load balancing, configuration,
formal systems for diagrammatic modeling and visual languages
and the corresponding graph systems are presented in
\cite{%
sim-viz-resource-alloc-control,%
visual-config-representation,%
logical-reasoning-with-diagrams,%
graph-transform-visual-languages,%
diagramatic-formal-system-euclidean,bpelse}.
They all define some key concepts that
are relevant to our visualization mechanisms within {\gipsy}
and its corresponding General Manager Tier (GMT) \cite{ji-yi-mcthesis-2011}.

We propose to build upon those works to represent the nested evidence,
crime scene as a 2D or even 3D {\dfg}, and the reconstructed events flow upon
evaluation.
Such a feature is projected in the near future to support
the previous work
on intensional forensic computing,
evidence modeling and encoding, and 
{\flucid}~\cite{%
flucid-imf08,%
flucid-blackmail-hsc09,%
flucid-raid2010,%
self-forensics-through-case-studies}
and {\marfl}~\cite{marfl-context-secasa08,marf-into-flucid-cisse08}
(where the intensional hybrid programming languages are being realized within the {\gipsy}
platform to investigate the languages' properties and test the run-time aspects thereof)
in order to aid investigator's tasks to build and evaluate
digital forensic cases.

\subsection*{Examples}

For that related work an conceptual example of a 2D DFG corresponding to a simple Lucid
program is in \xf{fig:nat42dfg}. The actual current rendering of such graphs is exemplified
in \xf{fig:ilucid-dfg-dot-yimin} from Ding \cite{yimin04} in the {\gipsy} environment.

In \xf{fig:nested-context-concept} is the
conceptual hierarchical nesting of the evidential statement $es$ context elements,
such as observation sequences $os$, their individual observations $o$ (consisting
of the properties being observed $(P,min,max,w,t)$, details of which are discussed
in the referenced related works). These 2D conceptual visualizations are proposed
to be renderable at least in 2D or in 3D via an interactive interface to allow modeling complex
crime scenes and multidimensional evidence on demand. The end result could look like
something expanding or ``cutting out'' nodes or complex-type results conceptually
exemplified in \xf{fig:nat42dfg-3d}

\begin{figure}[htpb]
	\includegraphics[width=\columnwidth]{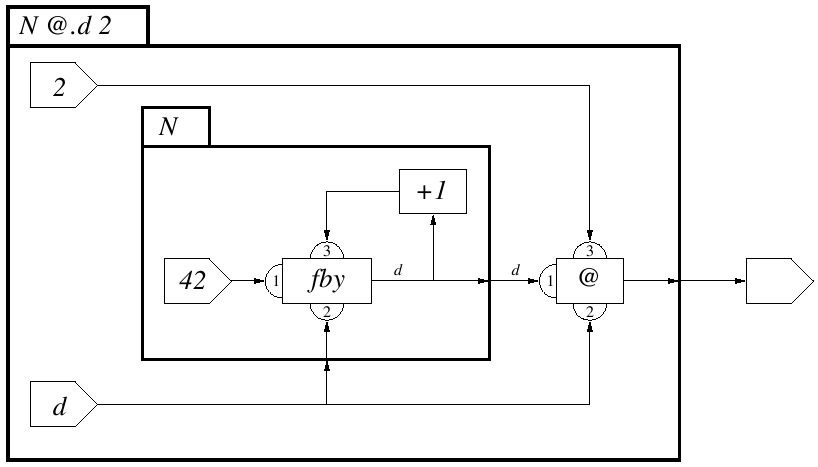}
	\caption{Canonical Example of a 2D Data Flow Graph-Based Program}
	\label{fig:nat42dfg}
\end{figure}

\begin{figure}[htpb]
	\includegraphics[width=\columnwidth]{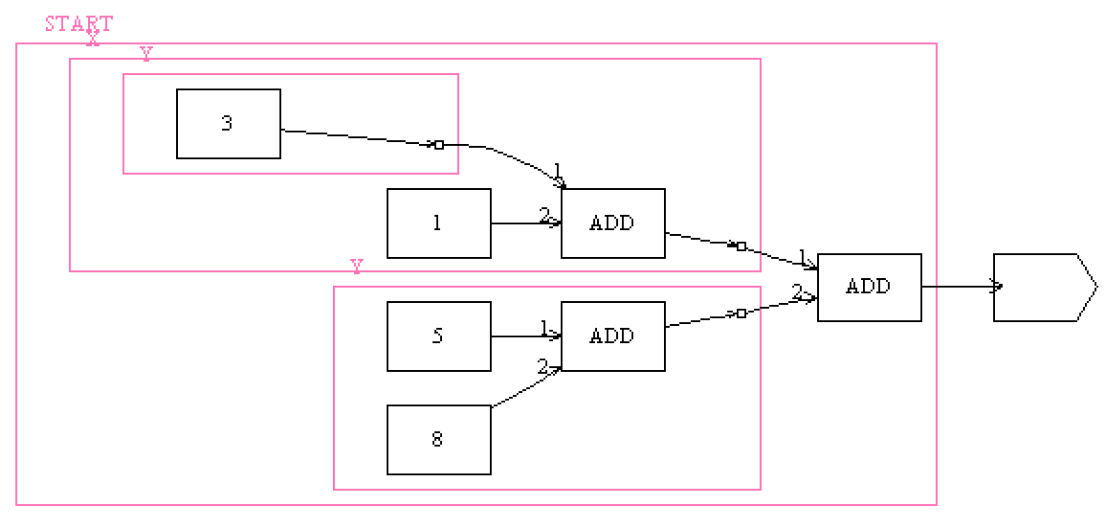}
	\caption{Example of an Actual Rendered 2D Data Flow Graph-Based Program with Graphviz \cite{yimin04}}
	\label{fig:ilucid-dfg-dot-yimin}
\end{figure}

\begin{figure}[htpb]
	\includegraphics[width=\columnwidth]{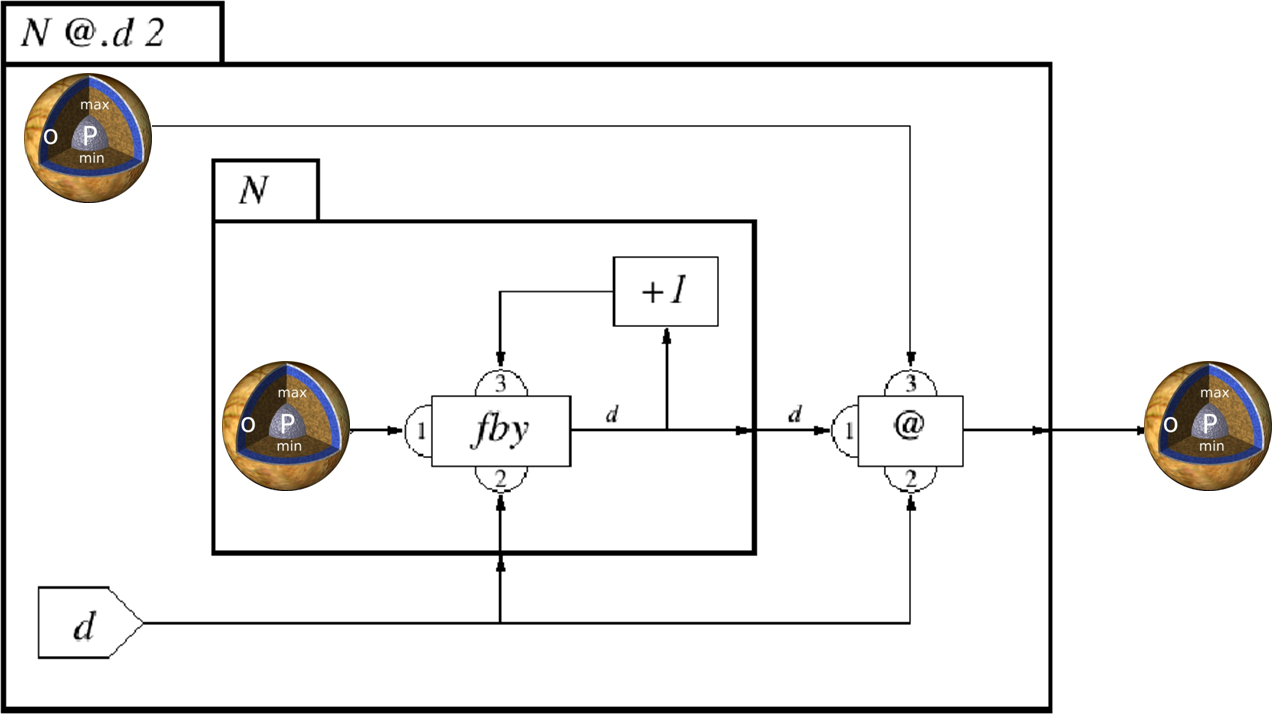}
	\caption{Modified Example of a 2D Data Flow Graph-based Program with 3D Elements.
		{Cutout image credit is that of Europa found on Wikipedia
		\url{http://en.wikipedia.org/wiki/File:PIA01130_Interior_of_Europa.jpg} from NASA}}
	\label{fig:nat42dfg-3d}
\end{figure}

\begin{figure}[htpb]
	\includegraphics[width=\columnwidth]{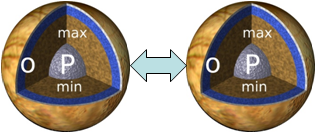}
	\caption{Conceptual Example of a 3D Observation Node.
		{Cutout image credit is that of Europa found on Wikipedia
		\url{http://en.wikipedia.org/wiki/File:PIA01130_Interior_of_Europa.jpg} from NASA}}
	\label{fig:observation-3d}
\end{figure}

\begin{figure}[htpb]
	\includegraphics[width=\columnwidth]{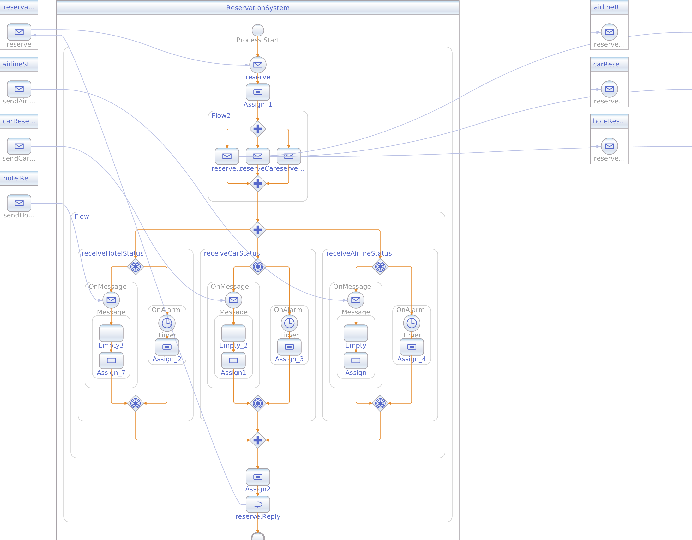}
	\caption{Example of a BPEL Graph with Asynchronous Flows \cite{bpelse}}
	\label{fig:bpel-example}
\end{figure}

\begin{figure}[htpb]
	\centering
	\includegraphics[width=\columnwidth]{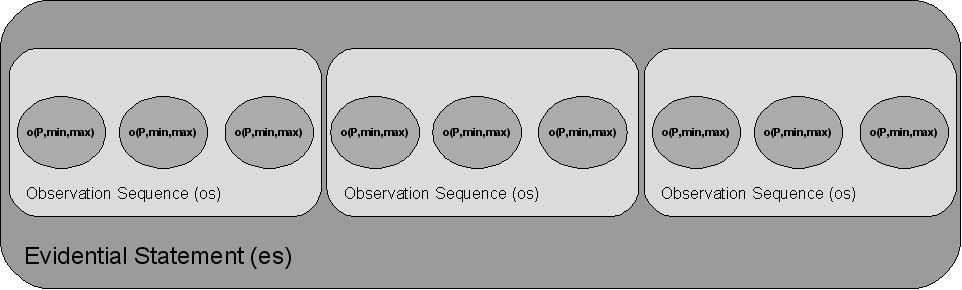}
	\caption{Nested Context Hierarchy Example for Cyberforensic Investigation~\cite{flucid-imf08}}
	\label{fig:nested-context-concept}
\end{figure}

\section{Visualization of {\flucid}}
\label{sect:3d-viz}

\subsection{3 Dimensions?}

The need to represent visually forensic cases, evidence,
and other specification components is obvious for usability and
other issues. Placing it in 3D helps to structure the ``program'' (specification)
and the case in 3D space can help arrange and structure the case in a virtual
environment better with the evidence items encapsulated in
3D balls like Russian dolls, and can be navigated in depth
to any level of detail via clicking (cf. \xf{fig:observation-3d}).

The depth and complexity of operational semantics
and demand-driven (eductive) execution
model are better represented and comprehended visually in 3D
especially when doing event reconstruction.
Ding's implementation allows navigation from a graph to a graph
by expanding more complex nodes to their definitions, e.g. more
elaborate operators such {\em whenever} (\api{wvr}) or {\em advances upon}
(\api{upon}).

\subsection{Requirements Summary}

Some immediate requirements to realize the envisioned
{\dfg} visualization of {\flucid} programs and their
evaluation:

\begin{itemize}
\item
Visualization of the hierarchical evidential statements (potentially
deeply nested context), cf. \xf{fig:nested-context-concept}.

\item
Placement of hybrid intensional-imperative nodes into the {\dfg}s
such as mixing {\java} and {\lucid} program fragments.
The {\gipsy} research and development group's
previous works did not deal with the way on how to augment the
\api{DFGAnalyzer} and \api{DFGGenerator} of Ding to support
hybrid GIPSY programs. This can be addressed
by adding an unexpandable imperative DFG node to the graph.
To make it more useful, i.e. expandable and so it's
possible to generate the GIPSY code off it or reverse it back.
The newer versions of {\graphviz} also have new support features
that are more usable for our needs at the present. Additionally,
with the advent of {\jooip} \cite{aihuawu09} the Java 5 ASTs
are available made available along with embedded {\lucid}
fragments that can be tapped into when generating the \tool{dot}
code's AST.

\item
Java-based wrapper for the DFG Editor of Yimin Ding \cite{yimin04}
to enable its native use withing Java-based {\gipsy} and plug-in
IDE environments like Eclipse.

\end{itemize}

\subsection{Selection of the Language and Tools}

One of the goals of this work is to find the optimal technique,
with soundness and completeness and formal specifications
ease of implementation and usability; thus we'd like
to solicit opinions and insights of this work in
selecting the technique or a combination of techniques,
which seems a more plausible outcome.

The current design allows any of the implementation to be
chosen or a combination of them.

\subsubsection*{{\graphviz}}

First, the most obvious is Ding's \cite{yimin04} basic DFG implementation
within {\gipsy} as it is already part of the project and done for the
two predecessor Lucid dialects. Additionally, the moder version of
{\graphviz} now also has integration with Eclipse \cite{eclipse}, so {\gipsy'}s
IDE -- {\ripe} (Run-time Interactive Programming Environment) -- may very
well be the an Eclipse-based plug-in.

\subsubsection*{{\puredata}}

Puckette came up with the {\puredata} \cite{puredata} language and its commercial
offshoots, which also employ DFG-like programming with boxes and
inlets and outlets of any data types graphically collected and allowing
sub-graphs and external implementations of inlets in procedural
languages. Puckette's original design was targetting signal processing
for electronic music and video processing and production for interactive artistic
and performative processes but has since outgrown that notion.
The {\puredata} externals allow deeper media visualizations in
OpenGL, video, etc. thereby potentially enhancing the whole
aspect of the process significantly.

\subsubsection*{BPEL}

The BPEL (Business Process Execution Language) and its visual
realization within NetBeans \cite{netbeans-671,netbeans} for SOA (service-orient architectures)
and web services is another good model for inspiration
\cite{koenig-ws-bpel-2007,ws-bpel-11} that has recently
undergone a lot of research and development, including flows, picking
structures, and faults, parallel/asynchronous and sequential activities.
More importantly, BPEL notations have a backing formalizm modeled
upon based on Petri nets (see e.g. visual BPEL graph in BPEL Designer (first came with the NetBeans IDE)
in \xf{fig:bpel-example} that illustrates two flows and 3 parallel
activities in each flow as well asynchrony and timeouts modeling.
This specification actually translates to an executable Java web services code).

\section{Conclusion}

With the goal to have a visual {\dfg}-based tool to model
{\flucid} case specification we deliberate on the possible
choice of the languages and paradigms within today's technologies
and their practicality and attempt to build upon previous
sound work in this area. Main choices so far identified
include Ding-derived {\graphviz}-based implementation,
{\puredata}-based, or BPEL-like. All languages are more or
less industry standards and have some formal backings;
the ones that don't may require additional work on to
formally specify their semantics and prove correctness
and sounds of translation to and from {\flucid}.

The main problem with {\puredata} and {\graphviz}'es \tool{dot}
is that their languages do not have formal semantics specified
only some semantic notes and lexical and grammatical structures
(e.g. see \tool{dot}'s \cite{dot-language}). If we use any
and all of these, we will have to provide translation rules
and their semantics and equivalence to the original {\flucid}
specification similarly as it is e.g was done by Jarraya for the UML 2.0/SysML state/activity diagrams
and probabilities in \cite{vv-uml-sysml-syseng-designs} when
translating to PRISM or equivalently for {\flucid} to PRISM
translation to do model-checking.

Thus, this work at this stage is to solicit comments
and recommendations on the proposed choices for the task.
Given the author's some familiarity with all three languages,
the final choice may result being and intermediate form
or a collection of forms mutually translatable.

\section*{Acknowledgments}

This work was supported in part by NSERC and
the Faculty of Engineering and Computer Science, Concordia University,
Montreal, Canada.

\bibliographystyle{plain}
\bibliography{gipsy-flucid-dfg}

\printindex

\end{document}